\documentclass[10pt,twocolumn,secnumarabic,amssymb,nobibnotes,aps,prx,superscriptaddress]{revtex4-2}

\usepackage{amsmath}
\usepackage{amsfonts}
\usepackage{amssymb}
\usepackage{graphicx}
\usepackage{hyperref}
\usepackage{longtable}
\usepackage{dcolumn}
\usepackage{bm}
\usepackage{epstopdf}
\usepackage{datetime}
\longdate
\usepackage{color}

\begin{document}

\title{Capillarity Reveals the Role of Capsid Geometry in HIV Nuclear Translocation}

\author{Alex~W.~Brown}
\thanks{These authors contributed equally}
\affiliation{Department of Physics, Durham University, Durham, DH1 3LE, U.K.}

\author{Sami~C.~Al-Izzi}
\thanks{These authors contributed equally}
\affiliation{School of Physics, UNSW, 2052 Sydney, NSW, Australia.}
\affiliation{ARC Centre of Excellence for the Mathematical Analysis of Cellular Systems, UNSW Node, Sydney, NSW 2052, Australia.}

\author{Jack~L.~Parker}
\affiliation{Department of Physics, Durham University, Durham, DH1 3LE, U.K.}

\author{Sophie~Hertel}
\affiliation{DTU Bioengineering, Department of Biotechnology and Biomedicine, Danmarks Tekniske Universitet, Søltofts Plads, Building 227 , 2800 Kgs.~Lyngby, Denmark.}

\author{David~A.~Jacques}
\affiliation{EMBL Australia Node in Single Molecule Science, School of Biomedical Sciences, UNSW, 2052 Sydney, NSW, Australia.}

\author{Halim~Kusumaatmaja}
\email{halim.kusumaatmaja@ed.ac.uk}
\affiliation{Department of Physics, Durham University, Durham, DH1 3LE, U.K.}
\affiliation{Institute for Multiscale Thermofluids, School of Engineering, University of Edinburgh, Edinburgh EH9 3FB, U.K.}

\author{Richard~G.~Morris}
\email{r.g.morris@unsw.edu.au}
\affiliation{School of Physics, UNSW, 2052 Sydney, NSW, Australia.}
\affiliation{EMBL Australia Node in Single Molecule Science, School of Biomedical Sciences, UNSW, 2052 Sydney, NSW, Australia.}
\affiliation{ARC Centre of Excellence for the Mathematical Analysis of Cellular Systems, UNSW Node, Sydney, NSW 2052, Australia.}

\begin{abstract}
    The protective capsid encasing the genetic material of Human Immunodeficiency Virus (HIV) has been shown to traverse the nuclear pore complex (NPC) intact, despite exceeding the passive diffusion threshold by over three orders of magnitude.
    This remarkable feat is attributed to the properties of the capsid surface, which confer solubility within the NPC's phase-separated, condensate-like barrier.
    In this context, we apply the classical framework of wetting and capillarity---integrating analytical methods with sharp- and diffuse-interface numerical simulations---to elucidate the physical underpinnings of HIV nuclear entry.
    Our analysis captures several key phenomena: the reorientation of incoming capsids due to torques arising from asymmetric capillary forces; the role of confinement in limiting capsid penetration depths; the classification of translocation mechanics according to changes in topology and interfacial area; and the influence of (spontaneous) rotational symmetry-breaking on energetics.
    These effects are all shown to depend critically on capsid geometry, arguing for a physical basis for HIV's characteristic capsid shape.

\end{abstract}

\maketitle

For ten of the eleven retrovirus genera, viral DNA is integrated into the host chromosome during cell division, when the nuclear envelope dissolves \cite{matreyek_viral_2013}. The remaining genus is the lentiviruses, whose representative type-species is Human Immunodeficiency Virus 1 (HIV-1) \cite{matreyek_viral_2013}. HIV and the other lentiviruses can infect non-dividing cells, and therefore must directly navigate the gatekeeper of the nucleus, the Nuclear Pore Complex (NPC).

The NPC is a large macromolecular assembly embedded in the double-walled nuclear envelope that is responsible for mediating nucleocytoplasmic transport via a diffusion barrier \cite{wente_nuclear_2010}. The barrier results from a family of proteins---the FG-Nucleoporins (FG-Nups)---which anchor in the walls of the NPC and project their intrinsically disordered regions into its central transport channel \cite{lin_structure_2019,denning_disorder_2003}. These regions are interspersed with small weakly-interacting phenylalanine-glycine (“FG”) motifs, which de-mix \textit{in vitro}, forming liquid condensate droplets \cite{schmidt_transport_2016}. In {\it situ}, the disordered regions of FG-Nups behave as if under Flory `good solvent' conditions, avoiding self-collapse and collectively (and dynamically) covering the transport channel \cite{yu_visualizing_2023}. The resulting milieu is estimated to limit passive diffusion for molecules $\gtrsim 30\,$kDa \cite{ribbeck_kinetic_2001,timney_simple_2016}.

Since the protective capsid that encases HIV's genetic material is $\sim50$ MDa [comprising 1000--1500 individual protein (CA) subunits \cite{briggs_stoichiometry_2004,li_image_2000,ganser_assembly_1999,gupta_critical_2023,mattei_structure_2016}] it was originally thought that this must disassemble to allow passage through the NPC \cite{hulme_identification_2015,francis_time-resolved_2016,francis_single_2018}. However, a wide variety of different assays \cite{burdick_hiv-1_2020,li_hiv-1_2021,zila_cone-shaped_2021,muller_hiv-1_2021,selyutina_nuclear_2020,dharan_nuclear_2020} all now indicate that HIV capsids transit the NPC and enter the nucleus intact.

\begin{figure*}
	\centering
	\includegraphics[width=0.99\textwidth]{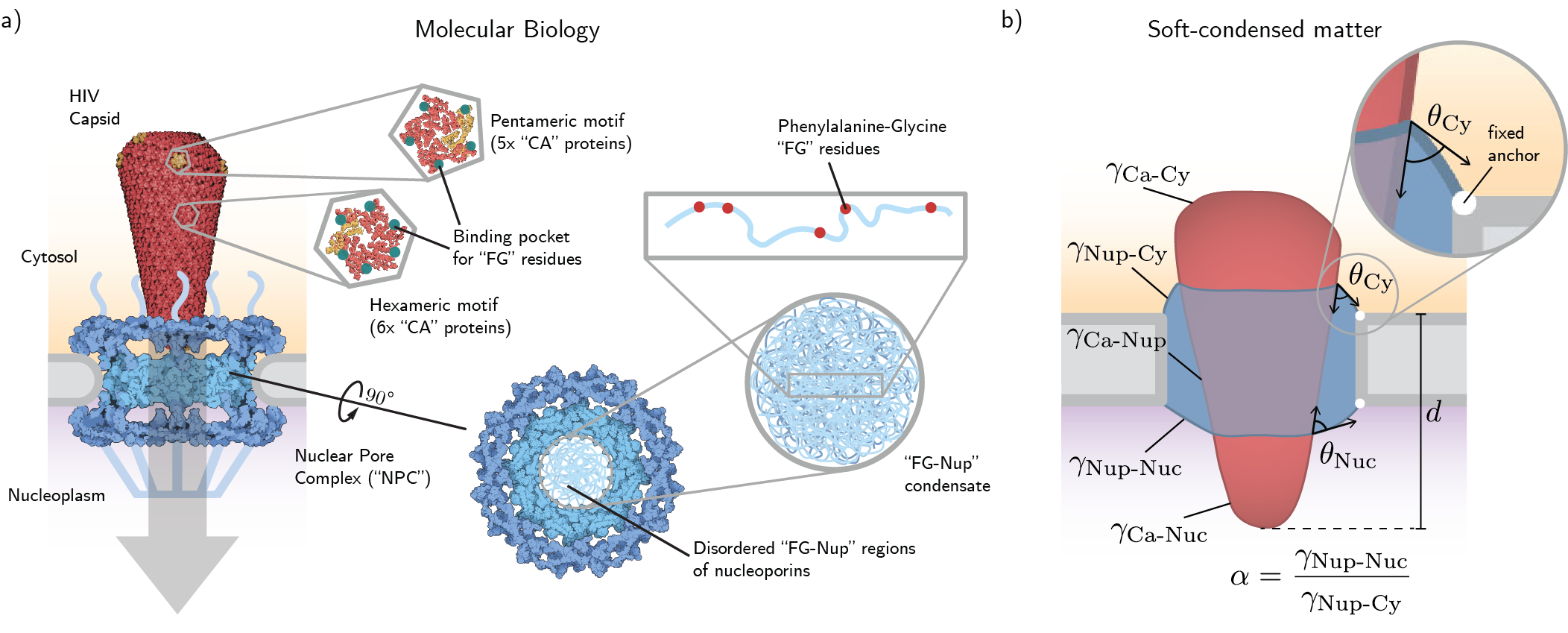}
	\caption
	{
		{\bf Casting HIV's transit through the NPC in terms of wetting and capillary forces}. The protective capsid that carries the viral genome of HIV has been shown to cross the NPC intact (panel a, \cite{zila_cone-shaped_2021}). This is remarkable, since the HIV capsid is over one thousand times larger than the limit for passive diffuse transport, which arises due to a barrier at the centre of the NPC. The diffusion barrier comprises hundreds of proteins (the FG-Nups) that are anchored in the walls of the central transport channel and whose disordered regions are interspersed with repeated weakly-interacting hydrophobic “FG” motifs, causing them to form a condensate-like milieu with ‘good solvent’ properties (panel a, \cite{yu_visualizing_2023}). The surface of the HIV capsid has been shown to be decorated by binding motifs \cite{dickson_hiv_2024} and other surface chemistry \cite{fu_governed_2025} that are complementary to the FG repeats, rendering it soluble in the diffusion barrier (panel a). Characterising these molecular interactions by per-unit-area free energy differences, the essential physics of HIV’s nuclear access is then captured by wetting and capillarity. The key variables are the ratio, $\alpha$, of FG-Nup:nucleoplasm and FG-Nup:cytosol interfacial energies, wetting angles at the cytoplasmic and nucleoplasmic interfaces---$\theta_\mathrm{Cy}$ and $\theta_\mathrm{Nuc}$, respectively---and the capsid geometry (panel b).
    }
	\label{fig:1}
\end{figure*}

That an object over 3 orders of magnitude larger than the NPC's limit for passive diffusion can breach the nuclear envelope is remarkable. It has since been attributed to \textit{i}) the structure of the individual CA protein subunits of the capsid, which incorporates a binding pocket that is capable of weakly interacting with FG-Nups (Fig.~\ref{fig:1}a)\cite{price_host_2014,lau_rapid_2021,dickson_hiv_2024}, and \textit{ii}) wider surface chemistry, where exposed molecular sidechains on the surface of CA proteins promote transient interactions with FG-Nups \cite{fu_governed_2025}. In other words, HIV capsids are soluble within the critical barrier that is supposed to keep them out.

This suggests that the way that capsids interact with FG-Nups
can be understood via a soft matter inspired approach. That is, replacing molecular details with interfacial free-energies, and invoking the phenomenology of wetting and capillarity \cite{gennes_capillarity_2003,pomeau_two_2006} (Fig.~\ref{fig:1}b). Here, we use theory and numerics to do precisely this.
In doing so, we reveal not only the physical basis of capsid nuclear access, but also the critical role played by geometry, addressing longstanding questions as to the function of HIV's characteristic shape, which is unique among the lentiviruses.

\section{Results}

\subsection{Capillary forces are controlled by two contact angles, a ratio of interfacial energies, and geometry}

\begin{figure*}[t]
	\centering
	\includegraphics[width=\textwidth]{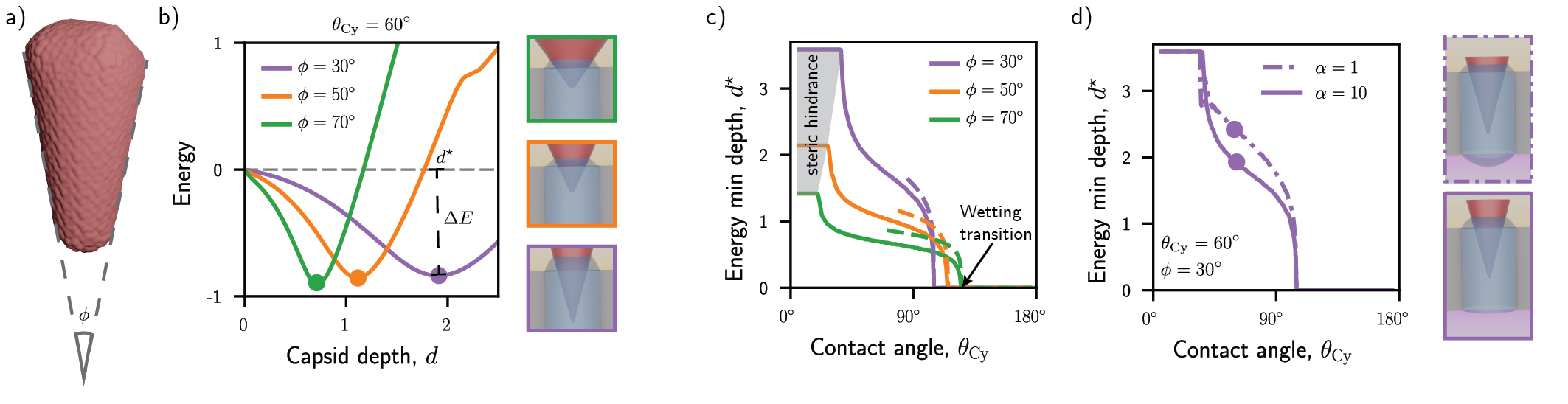}
	\caption
	{
        \textbf{Confinement, wetting and penetration}.
        Due to the confinement of the NPC's central channel, FG-Nups must be displaced for capsids to enter.
        The interplay between such confinement and the characteristic conelike geometry of capsids is captured by analysing cones with different angles, $\phi$, subtended at their tip (panel a).  
        For a given $\phi$ and (acute) $\theta_\text{Cy}$, the energy (\ref{eq:E}) has a non-trivial minimum as a function of penetration depth (panel b). This is because the energetic benefits of exchanging FG-Nup:cytosol and FG-Nup:capsid interfacial areas---and the resultant capillary forces---are eventually counterbalanced by increases in the total FG-Nup interfacial area due to increased FG-Nup displacement as the cone penetrates more deeply (panel b). 
        Since the amount of displaced FG-Nup increases less rapidly with penetration depth for cones with a smaller $\phi$, those capsids have a deeper energy-minimising penetration, $d^\star$ (panel b). 
        However, for similar reasons, the onset of the wetting transition occurs at smaller $\theta_\mathrm{Cy}$ for cones with smaller $\phi$, implying wetting over a smaller range of contact angles [panel c, solid lines represent numerics and dashed lines analytical approximation (Supplementary Material)]. 
        The effects of confinement and displacement are modulated by $\alpha$: smaller values of $\alpha$ permit displaced FG-Nup to be accommodated by the nucleoplasmic interface, and hence reduce the effects of $\phi$ (panel d).
        Unless otherwise specified, data was produced using sharp-interface numerics with $\alpha = 1$. 
    }
	\label{fig:2}
\end{figure*}

\begin{figure*}[t!]
	\centering
	\includegraphics[width=\textwidth]{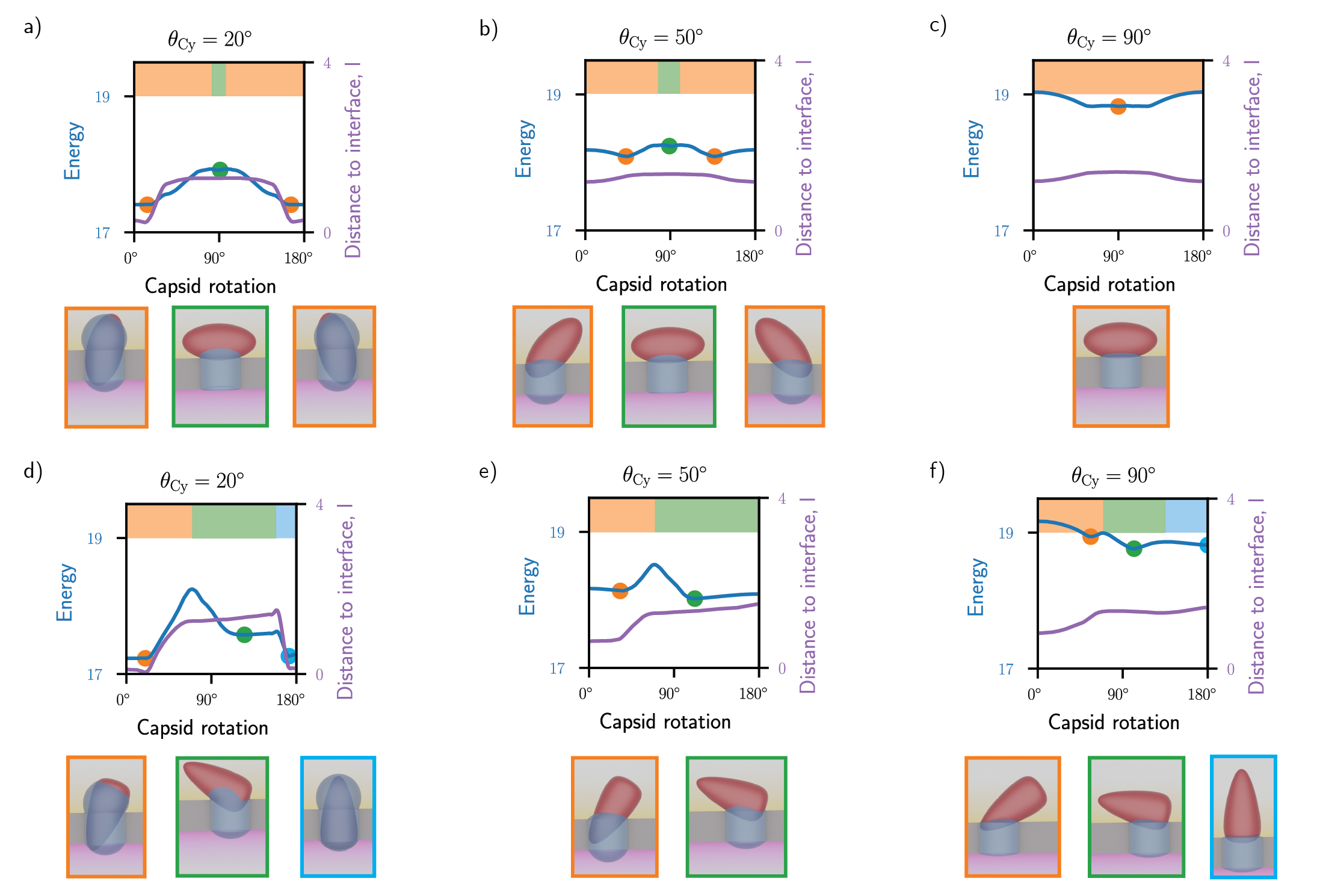}
	\caption
	{
        \textbf{Capillary-force driven reorientation}.
        Torques arising from asymmetric capillary forces can reorientate capsids arriving at the cytosolic side of the NPC. To capture this, we compute the initial energy minima of capsids penetrating the NPC that are fixed at different angles of rotation. Torque-free configurations are then given by the extrema of these landscapes.
        Such points (indicated by dots), their basins of attraction (indicated by overbars), the heights of the barriers that separate them, and the penetration depths to which they correspond, all depend on the shape of the capsid and the cytoplasmic contact angle, $\theta_\text{Cy}$ (see main text).
        Whilst increasing $\theta_\text{Cy}$ broadly reduces the heights of the barriers separating a given pair of minima, there are two notable differences between ellipsoidal and conelike shapes. First, conelike capsids have greater penetration depths, permitting contact with the nucleoplasmic interface at low $\theta_\text{Cy}$ (panels a \& d). Second, the local minimum of conelike capsids---with the tip pointing into the cytosol at an intermediate angle---not only has a much larger basin of attraction than their ellipsoidal counterpart, but becomes globally stable with increasing $\theta_\text{Cy}$ (panels d, e \& f).   
        Data was produced using sharp-interface numerics with $\alpha = 1$ and capsid shapes codified by (\ref{eq:3-parameter}) (Materials \& Methods). 
    }
	\label{fig:3}
\end{figure*}

Taking a classical approach to interfacial physics, we characterise the collective effects of transient interactions between individual molecular binding motifs by per-unit-area penalties to a free energy (Fig.~\ref{fig:1}).There are five relevant interfaces (Fig.~\ref{fig:1}b): between the capsid surface and the cytoplasm (“Ca-Cy”); between the capsid and the nucleoplasm (“Ca-Nuc”); between the capsid and FG-Nup condensate of the diffusion barrier (“Ca-Nup”); between the FG-Nups and the cytoplasm (“Nup-Cy”) and; between the FG-Nups and the nucleoplasm (“Nup-Nuc”).
However, due to the total surface area of the capsid being constant and once normalized by the FG-Nup:cytosol interfacial energy and the undeformed Nup-Cy area, the key variation in the energy associated with these five interfaces only depends on three degrees-of-freedom: two contact angles, $\theta_\text{Cy}$ and $\theta_\text{Nuc}$, at the ternary interfaces of capsid, FG-Nup, and cytosol / nucleoplasm, respectively, and the ratio $\alpha = \gamma_\text{Nup-Nuc} / \gamma_\text{Nup-Cy}$ of surface tensions associated with FG-Nup:nucleoplasm and FG-Nup:cytosol interfaces.
Relegating the details to the Supplementary Material, the dimensionless expression for the energy is  
\begin{equation}
    \begin{split}
        E &= A_\text{Nup-Cy} + \alpha\,A_\text{Nup-Nuc} + \cos\theta_\text{Cy}\,A_\text{Ca-Cy} \\
        &\quad + \alpha \cos\theta_\text{Nuc}\,A_\text{Ca-Nuc},
    \end{split}
    \label{eq:E}
\end{equation}
where the contact angles are marked on Fig.~\ref{fig:1}b, and a sub-scripted $A$ represents the area of a given interface. 
Since the FG-Nups are anchored in the walls of the NPC, with their disordered regions free to bend at negligible elastic cost, the FG-Nup:NPC-wall interface does not contribute to (\ref{eq:E}), either via changes in area or contact angle.

Despite the apparent simplicity of (\ref{eq:E}), its minimisation is contingent on boundary conditions in the form of both local and global geometrical constraints (Materials \& Methods).
Only a few analytical results are possible, and we resort to numerical techniques. 
We use two main approaches, a sharp interface approach, based on the well-known Surface Evolver software (\cite{brakke_surface_1992} and Materials \& Methods) and a diffuse interface, phase-field approach, using custom code (\cite{oktasendra_diffuse_2025} and Materials \& Methods).

\subsection{Penetration: more acute cone angles penetrate to greater depths but wet over a smaller range of contact angles}\label{sec:wetting}

The signature characteristic of HIV capsids is their cone-like shape.
The salient aspects, in terms of capillarity, are captured by conical capsids whose shape is characterised by a single control parameter: the angle subtended at its tip, $\phi$ (Fig.~\ref{fig:2}a).

For a given $\phi$ we fix the cytoplasmic contact angle, $\theta_{\text{Cy}}$, to be acute.
The energy (\ref{eq:E}) initially decreases with penetration depth, $d$---{\it i.e.}, capillary forces `suck' the capsid into the pore. This is because the size of the energetically-costly capsid:cytosol interface, $A_{\text{Nup-Cy}}$, decreases in favour of that of the less-costly capsid:FG-Nup interface, $A_{\text{Ca-Nup}}$ (Fig.~\ref{fig:2}b). This is similar to the physics of Pickering emulsions \cite{chevalier_review_2013}. However, a major difference is the role of confinement provided by the NPC as the conically-shaped capsid penetrates deeper.
Since increasingly large volumes of FG-Nup must be displaced per unit penetration, the total interfacial area, $A_{\text{Ca-Nup}}+ A_{\text{Nup-Cy}}$, also increases.
At some threshold, this outweighs the benefits of relative changes between $A_{\text{Ca-Nup}}$ and $A_{\text{Nup-Cy}}$, and $E$ begins to increase, resulting in a minimum-energy (equilibrium) penetration depth, $d^\star$ (Fig.~\ref{fig:2}b).
As the volume displaced at a given $d$ increases more rapidly for larger cone angles, $d^\star$ is larger for smaller $\phi$---\textit{i.e.}, the `thinner' a cone is, the deeper its equilibrium penetration (Fig.~\ref{fig:2}b).

To understand the role of the contact angle, we compute $d^\star$ as a function of $\theta_\text{Cy}$ for fixed $\phi$ (Fig.~\ref{fig:2}c).
Generically, this decreases monotonically from some maximum value---corresponding to the depth at which the cone sterically `hits' the top of the pore---until, at some critical $\theta^c_\text{Cy}$, the cone no longer wets (Fig.~\ref{fig:2}c).
For the same geometrical reasons as outlined above, however, $\theta_\text{Cy}^c$ decreases with $\phi$. That is, thinner shapes require more acute contact angles to wet (Fig.~\ref{fig:2}c).
This agrees with analytical calculations (Fig.~\ref{fig:2}c dashed lines); to leading order in $\phi$, the relevant second-order phase transition can be shown to satisfy $\cos\theta^c_\text{Cy}=-\phi/2$ (Supplementary Material). 

Since the displacement of the FG-Nup condensate is central to how different shapes interact with confinement, the equilibrium penetration depth can be controlled by changing $\alpha$, the ratio of nucleoplasmic and cytoplasmic interfacial energies.
As $\alpha$ decreases, the FG-Nup:nucleoplasm interface can be deformed for lower energetic cost, meaning that it `absorbs' more displaced volume and the cytosolic interface behaves increasingly as if there is no confinement (Fig.~\ref{fig:2}d).
In other words, cones with larger $\phi$ can penetrate as deeply as those with smaller $\phi$, so long as $\alpha$ is changed accordingly. More generally, we see that changing the contents of the nucleoplasm (and therefore $\gamma_\text{Nup-Nuc}$) can indirectly alter the penetration depth on the cytosolic side---via $\alpha$---{\it without} the capsid actually penetrating the second interface.    

\begin{figure*}[t]
	\centering
	\includegraphics[width=\textwidth]{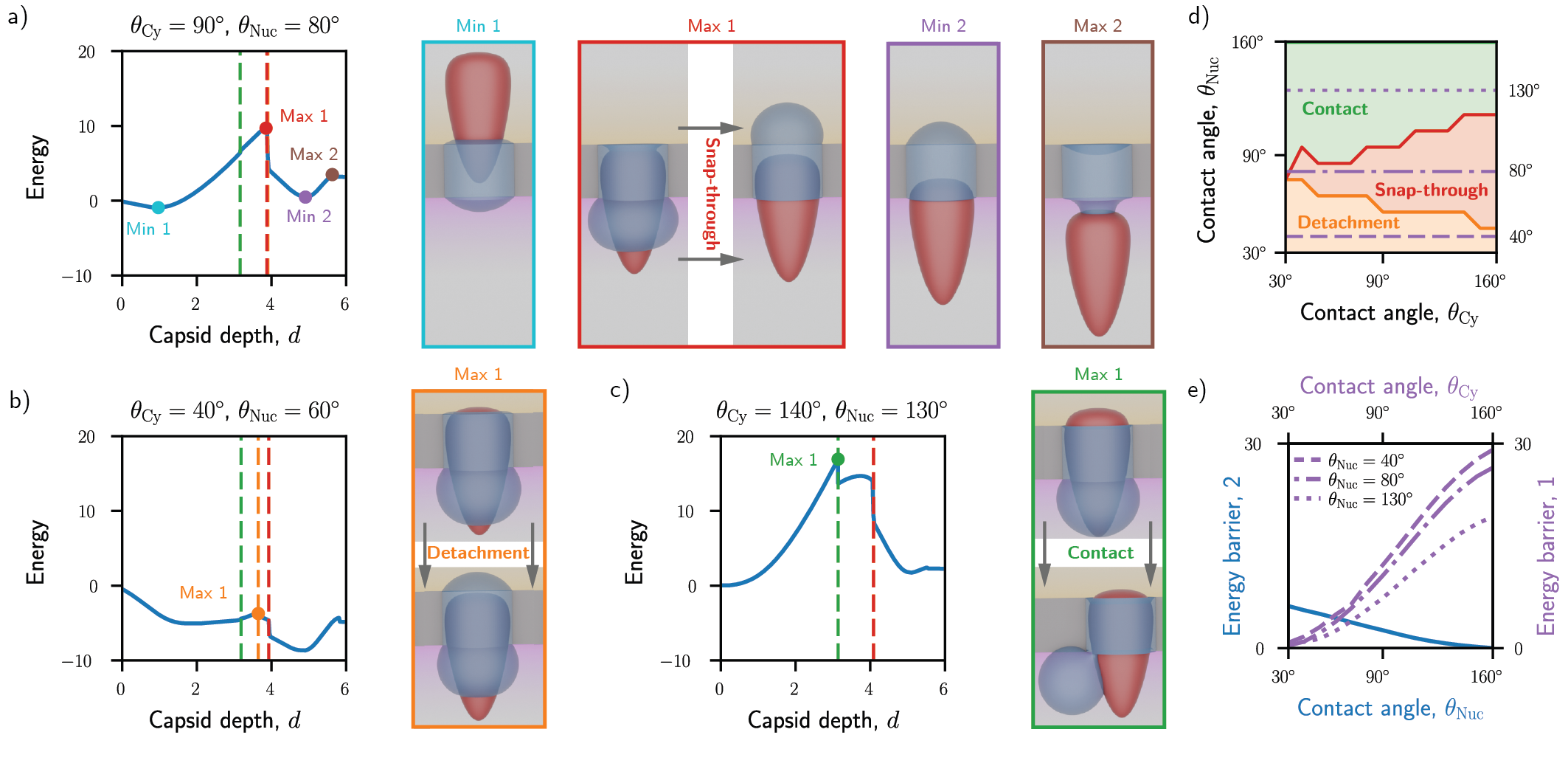}
	\caption
	{
        {\bf Classification and energetics of translocation: changes in topology and snap-through}.
        The translocation of conelike capsids through the NPC can be classified into one of three cases according to the mechanism underpinning the passage over the initial energy barriers.
        The first case corresponds to a snap-through transition, whereupon the cytosol:FG-Nup interface detaches from the capsid---changing the topology of the FG-Nup condensate---at the same time as the displacement of the FG-Nup condensate moves from the nucleoplasmic to cytosolic interfaces (panel a, red).
        The second corresponds to detachment only, so that the condensate changes topology but does not undergo a simultaneous snap-through (panel b, orange).
        The third case corresponds to making contact with the nucleoplasmic interface before the cytosolic one detaches (panel c, green).
        These three cases correspond to distinct regions of the configuration space parametrised by contact angles $\theta_\text{Cy}$ and $\theta_\text{Nuc}$ (panel d).
        Energetically, the different mechanisms broadly correspond to the rescaling of an otherwise monotonic dependence on $\theta_\text{Cy}$ (panel e, purple).
        The second energy barrier corresponds to detachment of the nucleoplasmic interface, irrespective of the mechanism underpinning the initial barrier.
        The energy of this barrier depends on $\theta_\text{Nuc}$ only (panel e, blue).
        For a large contact angle mismatch---{\it i.e.}, $\theta_\text{Cy} \lesssim 40^\circ$ and $\theta_\text{Nuc} \gtrsim 120^\circ$--- both energetic barriers are small and translocation likelihood is increased. Data was produced using sharp-interface numerics with $\alpha = 1$ and capsid shapes codified by (\ref{eq:3-parameter}) (Materials \& Methods).
    }
	\label{fig:4}
\end{figure*}

\subsection{Reorientation: conelike capsids are more likely to contact the nucleoplasmic interface at the cost of stabilising a shallow-penetration state.}\label{sec:reorientation}

Since capsids can arrive at cytosolic side of NPC with arbitrary orientations, they will typically experience asymmetric capillary forces.
These generate effective torques that act to reorientate capsids, with the details depending on both the shape of the capsid and cytosolic contact angle, $\theta_\text{Cy}$.
\par
We demonstrate such behaviour in the context of a three-parameter family of shapes [see Eq.~(\ref{eq:3-parameter}) in Materials \& Methods and Supplementary Material Sec. III.1]. 
This permits an interpolation between prolate ellipsoids---the simplest shape that can fit, sterically, through the channel---and conelike shapes (of the same surface area) that more closely represent the canonical HIV capsid. 
For a given shape and $\theta_\text{Cy}$, we compute the energy minimising configuration at different orientations---measured by the angle between the $x$-axis of the laboratory frame and the long axis of the capsid (for complete landscapes see Supplementary Material Sec. III.2).
The resulting profiles are characterized in terms of the torque-free states---\textit{i.e.}, those configurations that extremise the energy minima as a function of orientation angle--- in particular, their basins of attraction, the energy barriers between them, and penetration depths to which they correspond (Fig.~\ref{fig:3}).
\par
At low cytoplasmic contact angles ($\theta_\text{Cy} = 20^\circ$, Fig.~\ref{fig:3}a \& d), the impact of a more conical shape is that the global energy minimum corresponds to contact with, and potential penetration of, the nucleoplasmic interface (Fig.~\ref{fig:3}a \& d, orange). By contrast, the base of a conical shape permits greater stability and a larger basin of attraction for a low-penetration, almost-horizontal configuration, relative to its ellipsoidal counterpart (Fig.~\ref{fig:3}a \& d, green). There is also a base-first configuration that has a small basin of attraction and large penetration depth (Fig.~\ref{fig:3}d, blue). 
\par
At moderate cytoplasmic contact angles ($\theta_\text{Cy} = 50^\circ$, Fig.~\ref{fig:3}b \& e), both energetic barriers and penetration depths are generically lower, irrespective of shape (Fig.~\ref{fig:3}b \& e). However, whilst the minima for ellipsoidal capsids have the same relative stability, those for conelike capsids have switched; the tip-first configuration now corresponds to a local minimum, and the low-penetration configuration is now the global minimum (Fig.~\ref{fig:3}b \& e, orange and green). Notably, the base-first minima no longer appears (Fig.~\ref{fig:3}e). 
\par
At high cytoplasmic contact angles ($\theta_\text{Cy} = 90^\circ$, Fig.~\ref{fig:3}c \& f), there is only one global minimum for ellipsoidal capsids---the horizontal configuration. For conelike capsids we see the reappearance of the base-first configuration. Importantly, there is very little penetration with the capsid just sitting atop the pore in all cases.
\par

Taken together, we can draw two salient points. First, as in the previous section, longer, thinner conelike shapes at low cytoplasmic contact angles result in greater penetration depths, including reaching the nucleoplasmic interface ({\it e.g.}, Fig.~\ref{fig:3}a \& b, orange). However, this comes at `cost': they also stabilise a low-penetration configuration, where the capsid sits atop the NPC with its tip pointing into the cytosol at a wide angle (\textit{e.g.}, Fig.~\ref{fig:3}e, green). Since penetration depths that make contact with (or are within thermal fluctuations of) the nucleoplasmic interface are a likely pre-cursor to translocation, we therefore expect that a significant proportion of capsids will be found `trapped' in this local minimum.
Close inspection of recent \textit{in situ} cryo-EM experiments appear to corroborate this prediction \cite{kreysing_passage_2025,hou_hiv-1_2025}. Second, the base-first configuration is less favourable for translocation both in terms of penetration depth and basin of attraction.
   
%
\begin{figure*}[t!]
	\centering
	\includegraphics[width=\textwidth]{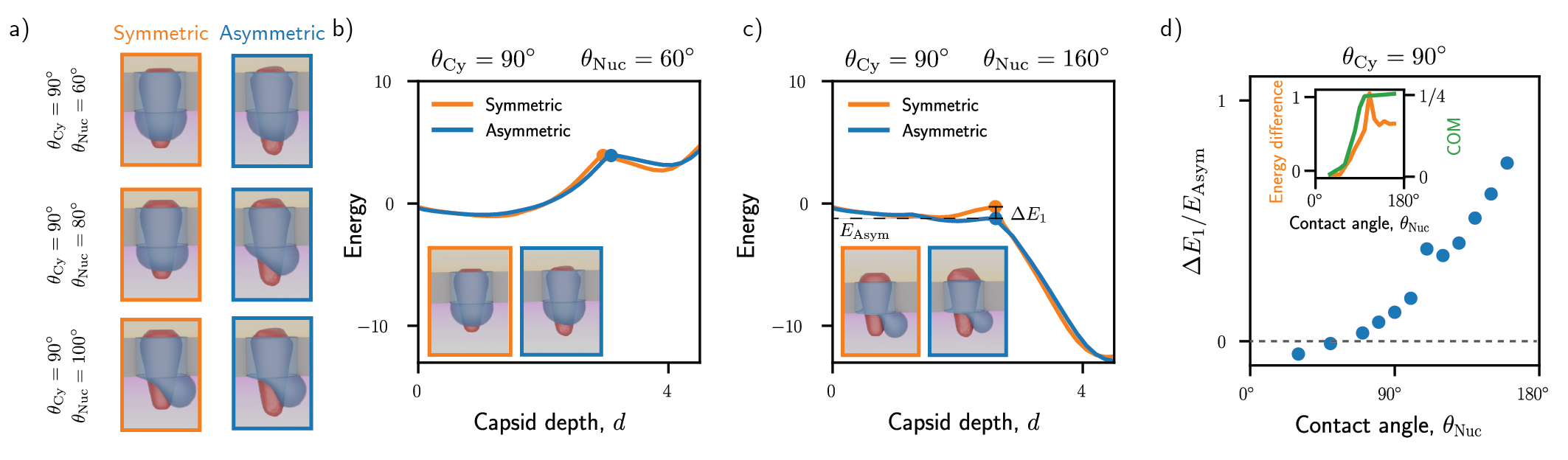}
	\caption
	{{\bf Symmetry-breaking of FG-Nups displaced into the nucleoplasm}.
    As rotationally-symmetric capsids approach the first energetic barrier to translocation, the displaced FG-Nup condensate can spontaneously break rotational symmetry, resulting in an asymmetric `bulge'.
    This occurs above a threshold $\theta_\text{Nuc}$, beyond which the size of the bulge increases with $\theta_\text{Nuc}$ (panel a, left). For `real' capsid shapes (\cite{ni_structure_2021} and Materials \& Methods) that also have asymmetry about their long axis, the symmetry-breaking of displaced FG-Nup is no longer spontaneously broken, and occurs for even small $\theta_\text{Nuc}$ (panel a, right top).
    Due to the energetic cost of a bulging FG-Nup condensate, the first barrier for symmetric capsids is (marginally) lower than that for `real' capsids for small $\theta_\text{Nuc}$ (panel b).
    Conversely, for large $\theta_\text{Nuc}$, the shape of real capsids can better accommodate the asymmetry of the FG-Nup condensate, resulting in smaller bulging as compared to symmetric capsids (panel a, right, middle and bottom).
    The height of the first barrier for symmetric capsids in this case is therefore higher than that for `real' capsids (panel c).
    Generically, the difference between the heights of the barriers for symmetric and `real' capsids (relative to the symmetric barrier height) increases as a function of $\theta_\text{Nuc}$ (panel d).
    Data was produced using diffuse-interface numerics with $\alpha = 1$ and capsid shapes derived from published data (Materials \& Methods).
    }
	\label{fig:5}
\end{figure*}

\subsection{Translocation: energy barriers can be classified according to topology and interfacial area}\label{sec:translocation1}

To better understand how capsids can fully translocate the NPC, we take vertically orientated conelilke capsids and compute the energy minimising configuration at all depths until the capsid has passed through the pore.
The resulting landscape is complex and depends on both cytoplasmic and nucleoplasmic contact angles.
\par
Apart from the initial and final configurations---\textit{i.e.}, when the capsid is fully embedded in the cytosol and nucleoplasm, respectively--- there are generically two stable local energy minima ({\it e.g.}, Fig.~\ref{fig:4}a).
These correspond to the capsid penetrating the cytoplasmic interface, but the not the nucleoplasmic one, and vice-versa.
\par
In between the two minima is a first energy barrier.
This can be classified according to one of three cases based on the mechanism that gives rise to the sudden drop in energy occurring immediately after the peak.
In the first case, the cytosolic interface detaches from the capsid---changing the global topology of the condensate from genus 1 to genus 0---at the same time as increasing its area. The area of the nucleoplasmic interface shrinks correspondingly, so that there is a sharp shift in displaced volume from the nucleoplasmic to cytosolic sides. 
We label this `snap-through' (Fig.~\ref{fig:4}a, red).
The second case is very similar, but with the barrier peak corresponding to detachment  ({\it i.e.}, the topological change) only, such that the sharp change in displaced volume occurs at deeper penetration depths.
We refer to this as `detachment' (Fig.~\ref{fig:4}b, orange).
The third case also corresponds to a topological change, but this time from genus 0 to genus 1, as the capsid first makes contact with the nucleoplasmic interface.
This we call `contact' (Fig.~\ref{fig:4}c, green).
\par
Between the second minima and full embedding in the nucleoplasm, there is a small second energy barrier.
This is associated with `pulling' the capsid out of the nucleoplasmic side ({\it e.g.}, Fig.~\ref{fig:4}a, purple).
In principle, this also happens on the cytosolic side, but we only consider translocation into the nucleus, and not the reverse.

\subsection{Translocation: contact angle mismatch reduces free energy barriers to promote translocation}\label{sec:translocation2}

The different classifications of the first energy barrier occur at different combinations of $\theta_\text{Cy}$ and $\theta_\text{Nuc}$ (Fig.~\ref{fig:4}d, see also Supplementary Material Sec. III.3).
At the lowest values of $\theta_\text{Cy}$, barriers correspond to either `detachment' (small $\theta_\text{Nuc}$) or `contact' (large $\theta_\text{Nuc}$).
As $\theta_\text{Cy}$ is increased, these two regimes are separated (at moderate $\theta_\text{Nuc}$) by a small regime of `snap-through'.
At the highest values of $\theta_\text{Cy}$, `snap-through' is the predominant mechanism, occurring at most values of $\theta_\text{Nuc}$, with `detachment' and `contact' only occurring at small and large $\theta_\text{Nuc}$, respectively.
\par
This mechanism-based structure broadly translates to the barrier height (Fig.~\ref{fig:4}d \& e, purple).
Whilst, generically, the height of the first barrier depends on $\theta_\text{Cy}$, this dependence scales with $\theta_\text{Nuc}$.
That is, for a given $\theta_\text{Cy}$, the `contact' mechanism has the lowest barrier height, followed by the `snap-through' mechanism, with `detachment' being the most energetically costly.
By contrast, the height of the second barrier only depends on $\theta_\text{Nuc}$ and not $\theta_\text{Cy}$ (Fig.~\ref{fig:4}e, blue).

\par
Of note, in the `contact' regime, where there is a large mismatch between contact angles---{\it e.g.}, $\theta_\text{Cy} \lesssim 40^\circ$ and $\theta_\text{Nuc}\gtrsim 120^\circ$---both barrier heights can be simultaneously made small. If they are reduced to a scale comparable to thermal fluctuations this would therefore facilitate full capsid translocation (Fig.~\ref{fig:4}e). For instance, taking a typical surface tension of $\gamma = 1 \mu$N/m for biomolecular condensates \cite{law_flicker_2023} and NPC diameter of 65 nm, $k_B T \sim 4$ in the unit of Fig.~\ref{fig:4}e, which suggests thermal fluctuations are sufficient to overcome the energy barriers for a significant fraction of the parameter space.

\begin{figure*}[t!]
	\centering
	\includegraphics[width=0.99\textwidth]{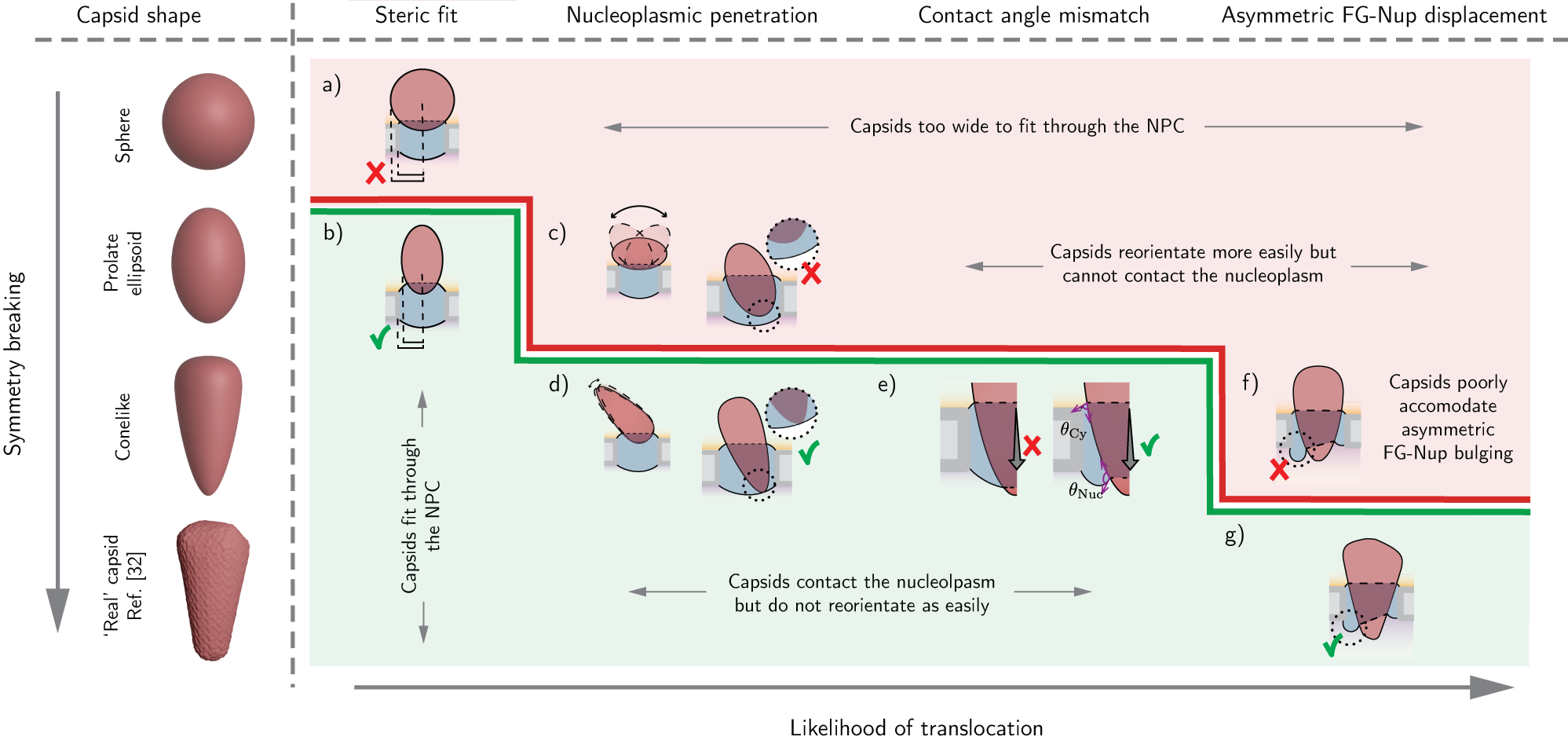}
	\caption
	{{\bf The effects of capsid geometry: `checkpoints' to translocation}. By setting the boundary conditions of (\ref{eq:E}), capsid geometry is central to the physics of a capsid's interactions with the NPC.
    This can be understood qualitatively in terms of breaking different symmetries in an ensemble of shapes of fixed surface area (Materials \& Methods). 
    Spherical capsids are too large to fit through the NPC's central channel (panel a). 
    Breaking symmetry along one axis results in prolate ellipsoids, which can fit through the pore (panel b). 
    However, whilst prolate ellipsoids can re-orientate easily, their maximum penetration depth does not typically result in contact with the nucleoplasmic interface, even at small $\theta_\text{Cy}$ (panel c). 
    Breaking the reflection symmetry of prolate ellipsoids results in conelike shapes. 
    Longer, thinner, cones penetrate more deeply so that contact with the nucleoplasmic interface can be made at the `cost' of stabilising a metastable state whereupon capsids point into the cytosol (panel d). 
    Once the nucleoplasmic interface has been breached, if $\theta_\text{Nuc}$ is large, the height of the largest energetic barrier is small and nucleoplasmic capillary forces effectively `suck' the capsid through the pore (panel e). 
    At such large $\theta_\text{Nuc}$, energy barriers can be further reduced by breaking a final symmetry around the capsid's long axis, thereby enabling a smaller `bulging' asymmetry of the displaced FG-Nup condensate (panels f \& g).
    }
	\label{fig:6}
\end{figure*}

\subsection{Translocation: breaking rotational symmetry around the capsid's long axis lowers energy barriers at high nucleoplasmic contact angles}
Despite both capsid and channel geometries being symmetric around the central axis, the FG-Nup condensate can spontaneously break this symmetry and form a `bulging' shape the instant the capsid penetrates the nucleoplasmic interface. This depends on the nucleoplasmic contact angle, first occurring at $\theta_\text{Nuc}\approx 80^\circ$ and becoming increasingly pronounced as $\theta_\text{Nuc}$ increases (Fig.~\ref{fig:5}a, orange and Fig.~\ref{fig:4}c, green).  

By contrast, `real' capsids---\textit{i.e.}, those derived from published data (Materials \& Methods)---typically have their rotational symmetry broken. In the context of the above, this results in two effects. First, the symmetry-breaking of the FG-Nup condensate no longer occurs spontaneously on contact with the nucleoplasmic interface: in order to accommodate an asymmetric capsid shape, the condensate must be asymmetric at all points during translocation. This means that, whilst the bulging effect on the nucleoplasmic side still becomes more prominent with increasing $\theta_\text{Nuc}$, there is no-longer a lower critical value below which condensates are symmetric (Fig.~\ref{fig:5}a, blue). Second, `real' asymmetric capsids are better able to accommodate the highly asymmetric bulging that occurs at larger values of $\theta_\text{Nuc}$. Since the direction in which a bulge occurs is dictated by the shape of the capsid (and not spontaneously chosen, as for symmetric capsids), the system is able to form bulges that are slightly smaller in size, for a given $\theta_\text{Nuc}$, than those of symmetric capsids.   

These heuristic observations are borne out by the heights of the relevant energetic barriers, which correlate with size of a bulge. At low $\theta_\text{Nuc}$, asymmetric capsids have a marginally higher (first) energy barrier due to a slight nucleoplasmic bulge, which does not occur for symmetric capsids (Fig.~\ref{fig:5}b). However, at high $\theta_\text{Nuc}$, the asymmetry of the capsid now leads to a lower (first) energy barrier, since the corresponding FG-Nup bulge is smaller than that formed spontaneously due to a symmetric capsid (Fig.~\ref{fig:5}c). Overall, the difference in the barrier heights of the two capsid shapes (divided by the height of the symmetric capsid) steadily increases as a function of $\theta_\text{Nuc}$ (Fig.~\ref{fig:5}d). The non-monotonicity around $\theta_\text{Nuc}\approx 110^\circ$ corresponds to the change in mechanism between `snap-through' and `contact'.

\section{Discussion}
We take a soft matter inspired approach to characterising viral capsids interacting with the NPC.
This captures phenomenology at timescales much longer than inter-molecular binding intervals and distances much greater than intermolecular spacings \cite{banani_biomolecular_2017}. 
It is, we argue, complementary to molecular dynamics \cite{hudait_hiv-1_2024,perez-segura_atomistic_2025}, offering both conceptual insight and efficient exploration of parameter space at the expense of molecular details.

The central physics is that of capillarity \cite{gouveia_capillary_2022}: a framework that lays bare the role of geometry. This is important because, despite pleomorphic shape variation due to stochastic self-assembly \cite{ganser_assembly_1999}, HIV capsids are broadly characterised by pear- or cone-like shapes: relatively unique geometries whose role has long been a source of intense speculation \cite{muller_nuclear_2022}.

In this context, it is instructive to revisit our results though the lens of symmetry (Fig.~\ref{fig:6}). For a representative, fixed surface area---capsids typically comprise circa.~1200 CA subunits \cite{briggs_stoichiometry_2004}---symmetrical, spherical capsids cannot fit through the NPC (Fig.~\ref{fig:6}a). This requires breaking symmetry along one axis to form a prolate ellipsoid (Fig.~\ref{fig:6}b). However, a prolate ellipsoid is unlikely to penetrate sufficiently deeply so as to make contact with the nucleoplasmic interface (Fig.~\ref{fig:6}c). This requires breaking the long-axis symmetry of prolate ellipsoids and forming a cone-like shape (Fig.~\ref{fig:6}d). Under mismatched contact angles, cones can experience almost barrier-free translocation (Fig.~\ref{fig:6}e). However, further breaking rotational symmetry can aid this process due to a smaller condensate bulging effect (Fig.~\ref{fig:6}f \& g). In other words, the more large-scale symmetries that are broken, the higher the likelihood of translocation.

The implication is that the assembly of HIV capsids may have evolved to produce shapes that are beneficial for translocation. A related question is whether this offers a potential basis for the rational design of capsids in biotechnological applications; transporting cargoes across the NPC is an important aspect of many emergent biotechnologies, with modified viral capsids already used in the delivery of tools for genetic modification \cite{kay_2001}.

Addressing such questions, however, is likely to require more nuanced considerations in several areas.
In NPC mechanics, these might include inner-ring dilation \cite{elosegui-artola_force_2017,andreu_mechanical_2022} and/or fracture \cite{kreysing_passage_2025}, as well as the compressibility of FG-Nup condensates.
Similarly, a more detailed understanding of how the molecular composition of the cytosol and nucleoplasm affect $\theta_\text{Cy}$ and $\theta_\text{Nuc}$, respectively, would lead to better quantification.
For example, Cleavage and Polyadenylation Factor 6 (CPSF6) is a predominantly nucleoplasmic protein that has been shown to bind to the non-specific CA protein site to which FG-Nups bind.
Increased expression of this protein is therefore expected to decrease the interfacial energy, $\gamma_\text{Ca-Nuc}$, and therefore increase the nucleoplasmic contact angle, $\theta_\text{Nuc}$.
It may also provide a screening effect for FG-repeats and hence alter $\gamma_\text{Nup-Nuc}$ and $\alpha$, and, as such, have a potentially profound effect on the energetics of nuclear entry.

\section{Materials \& Methods}\label{sec:materialsAndMethods}

\subsection{Setup} 
We use fixed dimensions for the NPC's central channel that are consistent with {\it in situ} cryo-EM images as a capsid passes through the pore \cite{zila_cone-shaped_2021}. Surfaces are assumed to be stiff and cannot bend, twist or stretch.
The inner channel is taken to be $65\,$nm in diameter and $50\,$nm in depth. To ensure that capsids that can fit through the pore, we set a maximum capsid diameter of $60\,$nm. Structural images suggest capsids do not exceed an aspect ratio of 2, so we consider maximum heights of $120\,$nm \cite{goryaynov_single-molecule_2011,zila_cone-shaped_2021}. These dimensions imply surface areas of approximately $1.9\times10^{4}\,$nm$^2$. This is in-line with estimates derived from the area of hexameric motifs of individual CA proteins. Assuming that there are approximately 250 hexamers, each with a diameter of $9\,$nm, this gives a total capsid surface area of $1.3\times10^{4}\,$nm$^2$ \cite{ganser-pornillos_structure_2007}. All lengths are non-dimensionalised relative to the inner channel radius.

We assume that the FG-Nup condensate is incompressible. Since HIV capsids occupy such a large volume fraction of the central channel during translocation, this helps us to qualitatively capture the phenomenology of FG-Nup displacement. Quantitatively, however, our results only provide a bound on, \textit{e.g.}, energetic barriers to translocation, since FG-Nups are likely to have some small but nonzero compressibility.

\subsection{Sharp interface: Surface Evolver} 

The numerical calculations in Figs.~\ref{fig:2}-\ref{fig:4} were carried out in Surface Evolver \cite{brakke_surface_1992}. This performs gradient descent for discrete triangulated surfaces via mean curvature flow. FG-Nup:cytoplasm and FG-Nup:nucleoplasm surfaces are pinned where they contact the wall of the NPC's channel, but are free to move where they contact the capsid, subject to fixed $\theta_\text{Cy}$ and $\theta_\text{Cy}$, respectively.

As capsids are treated as solid, we fix interfaces to lie on either a cone with tip-angle $\phi$ ({\it e.g.}, Fig.~\ref{fig:1}) or a shape defined by
\begin{equation}
    \left(\frac{x^2 +y ^2}{C^2}\right)\left(1-Sz\right) + \left(\frac{z}{h}\right)^2 = 1,
    \label{eq:3-parameter}
\end{equation}
({\it e.g.}, Figs.~\ref{fig:3} \& \ref{fig:4}). Here, $S$ controls the asymmetry of the capsid. Setting $ S=0 $ recovers the equation for an ellipsoid where $C$ and $h$ become the minor and major axes respectively. As we increase $S$ to increase the asymmetry of the capsid we alter $C$ to fix the surface area of the capsid which in turn fixes the surface energy of the capsid between symmetric and symmetry-broken configurations. In units of pore radius $(C,S,h)=(0.923,0.0,1.846)$ for ellipse and $(C,S,h)=(0.821, 21.0, 1.846)$ for the cone.

Energetics are quasi-static: they are obtained by placing a capsid at a given penetration depth (and/or angle of rotation) and then minimising the surface energies (and hence geometries). The process is then repeated to obtain  energy landscapes, the gradients of which correspond to capillary forces. The energy landscapes in Fig.~\ref{fig:3} are smoothed using a Savitzky-Golay filter (window size $25$ and polynomial order $4$) to allow for automated finding of the minima with minimal user input.

\subsection{Diffuse interface: phase-field} 

To model `real' HIV capsid shapes that are obtained from data and are not easily codified by a closed analytical form, we use a diffuse-interface method based on Ref.~\cite{oktasendra_diffuse_2025}. A lattice-based  phase field simulation is more expensive than the continuum modelling in Surface Evolver. However, with the addition of a frozen fluid scheme, the complex shape of the HIV capsid can be accommodated.

We employed a three-component phase field model with the following free-energy functional
\begin{multline}
        F = \int_{\Omega}\Big[ \frac{\kappa_1(\bf{x})}{2}C_1^2(1-C_1)^2 + \frac{\kappa_2}{2}C_2^2(1-C_2)^2 \\
        + \frac{\kappa_3}{2}C_3^2(1-C_3)^2 +\frac{\kappa_1'(\bf{x})}{2} (\nabla C_1)^2\\
        + \frac{\kappa_2'}{2} (\nabla C_2)^2+ \frac{\kappa_3'}{2} (\nabla C_3)^2\Big]\mathrm{d}V_\Omega.
\label{eq:free_energy_functional}
\end{multline}
Each component has a bulk concentration value of $C_i = 0$ when it is absent and $C_i = 1$ when it is present. We imposed the hard constraint, $C_1 + C_2 + C_3 = 1$. Since there is never a cytoplasm-nucleoplasm interface, we can use $C_1$ to represent the cytoplasm component above the halfway plane of the NPC and the nucleoplasm component below the halfway plane of the NPC. $C_2$ and $C_3$ represent the FG-nup and capsid components. 

Without loss of generality, upon crossing the interface from components $i$ to $j$, the concentrations vary smoothly as 
\begin{equation}
    C_i = \frac{1-\tanh{\frac{x}{2\xi}}}{2}, \;\;\;\;
    C_j = \frac{1+\tanh{\frac{x}{2\xi}}}{2},
\label{eq:tanh}
\end{equation}
where $\xi$ is the interface width, and $x$ is the signed distance from the interface. For a given interface width $\alpha$, $\kappa_i' = \xi^2 \kappa_i$, and $\kappa_1$, $\kappa_2$, $\kappa_3$ can be tuned to vary the interfacial tensions. The interfacial tension between components $i$ and $j$ is given by
\begin{equation}
\gamma_{ij} = \frac{\xi}{6} (\kappa_i + \kappa_j),
\end{equation}
and the FG-nup contact angle on the capsid is
\begin{equation}
\cos \theta = \frac{\gamma_{13}-\gamma_{23}}{\gamma_{12}}.
\end{equation}
By making $\kappa_1(\bf{x})$ position-dependent (whether $C_1$ represents the cytoplasm or nucleoplasm), we can specify $\theta_{\rm{Cy}}$ and $\theta_{\rm{Nuc}}$. 

The frozen fluid scheme proceeds in two steps, as detailed in Ref.~\cite{oktasendra_diffuse_2025}. The first step initializes the capsid shape with a suitable diffuse interface profile as given by Eq.~\eqref{eq:tanh}. The concentration distribution of the $C_3$ component is then fixed, and the FG-nup, cytoplasm, and nucleoplasm components are introduced to the simulation domain. To obtain the minimum energy configuration for a given capsid position, we employed the L-BFGS algorithm, a quasi-Newton method that approximates second-order curvature using information from recent steps.

The asymmetric HIV capsid shape used was taken from the PDB in a process explained in III.4. The symmetric capsid used in the comparison took the average profile of the asymmetric capsid in $z$ and rotated around the $z$-axis. Similar to our approach in III.3, to obtain the energy landscapes, the HIV capsids were placed within the pore at many locations on the $z$-axis, and the minimum energy configuration was obtained at each instance.

\subsection{Smooth density maps of the HIV capsid fullerene cone}

The atomic coordinates of the wild-type HIV capsid fullerene cone were obtained from cryo-electron microscopy density maps of the HIV capsid from Ni \textit{et al.} \cite{ni_structure_2021}.
The atomic coordinates were then converted into a smooth Gaussian density map using the molecular viewer VMD 1.9.4 \cite{humphrey_vmd_1996}. The Quicksurf algorithm \cite{meyer_fast_2012} was used to generate the density map, with a radius scale setting of 8, density isovalue of 4, grid spacing of 4 and medium surface quality. The structure was then rendered with VMD to STL format to obtain a triangulated mesh. Vertices in the interior of the capsid were removed, leaving only the exterior vertices of the volume map.

\section{Author Contributions}
SCA, SH, DAJ, HK and RGM conceived of the project. SCA performed analytical calculations. AWB performed numerical simulations. JLP developed custom phase-field code. AWB and SCA analysed simulation data. HK and RGM oversaw the project, with contributions from all authors. RGM, HK, AWB and SCA wrote the manuscript with critical feedback from all authors.

\section{Acknowledgments}
SCA and RGM acknowledge the EMBL Australia program and funding from the Australian Research Council Centre of Excellence for Mathematical Analysis of Cellular Systems (CE230100001). DAJ, HK and RGM acknowledge funding from the Australian Research Council (DP240103034). HK acknowledges funding from EPSRC EP/V034154/2. AWB acknowledges funding from EPSRC CDT on Molecular Sciences for Medicine EP/S022791/1. The authors acknowledge Dr.~Brian Ee, who helped obtain smooth density maps of HIV structures, and Dr.~Minkush Kansal and Prof.~Till B\"ocking for helpful comments and feedback.

\bibliographystyle{apsrev4-1}
\bibliography{references}

\clearpage
\onecolumngrid
\makeatletter 
\def\tagform@#1{\maketag@@@{(S\ignorespaces#1\unskip\@@italiccorr)}}
\makeatother
\graphicspath{{figures/}} 
\makeatletter
\makeatletter \renewcommand{\fnum@figure}
{\figurename~S\thefigure}
\makeatother
\def\eq#1{{Eq.~S\ref{#1}}}    
\def\fig#1{{Fig.~S\ref{#1}}}
\setcounter{figure}{0} 
\setcounter{equation}{0} 
\setcounter{section}{0} 
\section*{Supplemental Material for ``Capillarity Reveals the Role of Capsid Geometry in HIV Nuclear Translocation''}\label{sec:SI}
\begin{center}
    Alex~W.~Brown, Sami C.~Al-Izzi, Jack~L.~Parker, Sophie Hertel, David~A.~Jacques, Halim~Kusumaatmaja \& Richard~G.~Morris 
\end{center}

\section{Energetics}

As described in the main manuscript, there are five relevant interfaces: between the capsid surface and the cytoplasm (“Ca-Cy”); between the capsid and the nucleoplasm (“Ca-Nuc”); between the capsid and FG-Nup condensate of the diffusion barrier (“Ca-Nup”); between the FG-Nups and the cytoplasm (“Nup-Cy”) and; between the FG-Nups and the nucleoplasm (“Nup-Nuc”).  Assigning a per-unit-area energy (or equivalently, a surface tension) to each interface, the total variation in energy due to changes in the areas of those interfaces is then just
\begin{equation}
    E = \gamma_\text{Ca-Cy}A_\text{Ca-Cy} + \gamma_\text{Ca-Nuc}A_\text{Ca-Nuc} + \gamma_\text{Ca-Nup}A_\text{Ca-Nup} + \gamma_\text{Nup-Cy}A_\text{Nup-Cy}+ \gamma_\text{Nup-Nuc}A_\text{Nup-Nuc}.
    \label{eq:total_energy}
\end{equation}
However, we also have that
\begin{equation}
    A_\text{Ca-Cy} + A_\text{Ca-Nup} + A_\text{Ca-Nuc} = \text{Const.},
    \label{eq:area_constraint}
\end{equation}
since the total surface area of the capsid is fixed.  Substituting (\ref{eq:area_constraint}) into (\ref{eq:total_energy}) and dividing by the quantity $\pi R^2 \gamma_\text{Nup-Cy}$, where $R$ is the pore radius, gives rise to the dimensionless relation
\begin{equation}
    \tilde{E} = \frac{\left(\gamma_\text{Ca-Cy} - \gamma_\text{Ca-Nup}\right)}{\gamma_\text{Nup-Cy}} \tilde{A}_\text{Ca-Cy} + \frac{\left(\gamma_\text{Ca-Nuc}-\gamma_\text{Ca-Nup}\right)}{\gamma_\text{Nup-Cy}} \tilde{A}_\text{Ca-Nuc} + \tilde{A}_\text{Nup-Cy}+ \frac{\gamma_\text{Nup-Nuc}}{\gamma_\text{Nup-Cy}} \tilde{A}_\text{Nup-Nuc},
\end{equation}
where a factor of $\pi R^2 \gamma_\text{Nup-Cy}$ has been absorbed into the definition of $\tilde{E}$, on the left-hand side and all the areas are non-dimensionalised by the pore are $\pi R^2$. Writing $\alpha = \gamma_\text{Nup-Nuc} / \gamma_\text{Nup-Cy}$, we may twice invoke Young's equation \cite{doi_soft_2013} to write
\begin{equation}\label{eq:simplifiedEnergy}
    \tilde{E} = \cos\theta_\text{Cy}\, \tilde{A}_\text{Ca-Cy} + \alpha \cos\theta_\text{Nuc}\, \tilde{A}_\text{Ca-Nuc} + \tilde{A}_\text{Nup-Cy}+ \alpha\, \tilde{A}_\text{Nup-Nuc},
\end{equation}
which recovers the relation provided in the main text, with tilde's dropped.

\section{Analytical model for a cone and single interface}
\begin{figure}
\center\includegraphics[width=\textwidth]{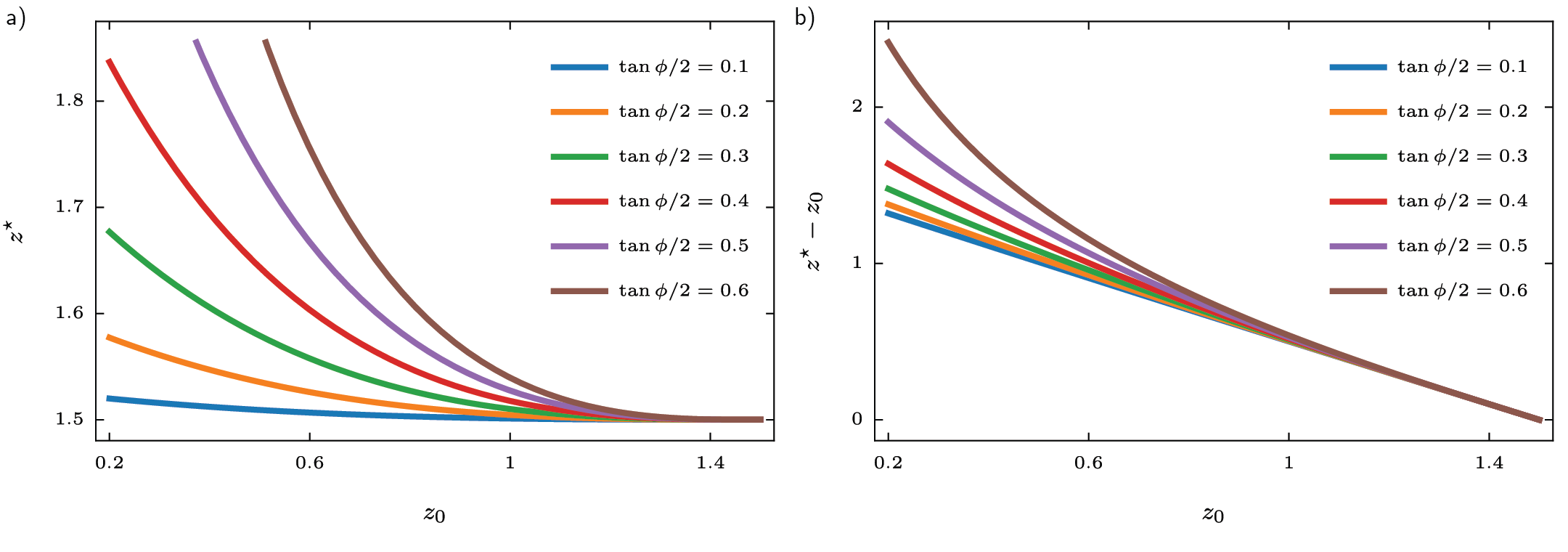}
\caption{\textbf{(a)} Level the gel raises to, $z^\star$, as a function of tip coordinate, $z_0$. \textbf{(b)} Height of the cone embedded, $z^\star - z_0$ as a function of tip coordinate, $z_0$.}
\label{fig:EmbeddingValuesSimple}
\end{figure}

\begin{figure}
\center\includegraphics[width=\textwidth]{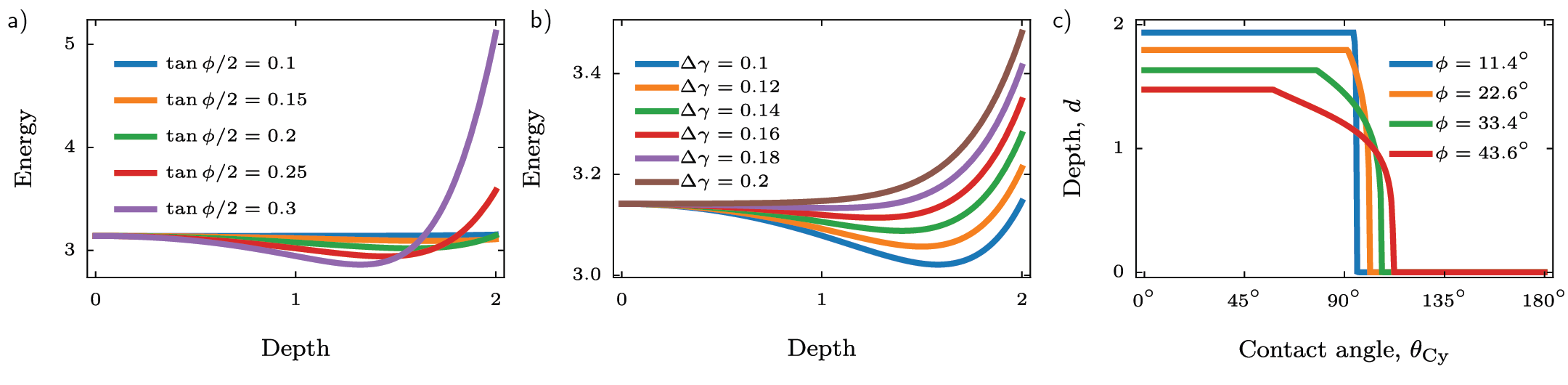}
\caption{\textbf{(a)} Free energy of the system as a function of capsid depth for a variety of cone angles, $\phi$, \textbf{(b)} Free energy as a function of depth for a variety of dimensionless surface tension differences, $\Delta\gamma$. \textbf{(c)} Insertion depth of the capsid in the minimum energy state as a function of contact angle, $\theta_\text{c}$ in degrees. This is plotted for three values of cone angle $\phi$.}
\label{fig:phaseDiagramsSimple}
\end{figure}

Here we analyse in detail a simple geometric model of a cone of angle $\phi$ where we consider only the cytoplasmic side of the pore; the total volume of the FG-Nup condensate is conserved and the FG-Nup:cytosol interface is taken to be a straight line. We can rewrite the energy, Eq.~\ref{eq:simplifiedEnergy}, as the following
\begin{equation}
    \tilde{E} = -\cos\theta_\text{Cy}\, A_\text{Ca-Nup} + A_\text{Nup-Cy} = \Delta \gamma A_\text{Ca-Nup} + A_\text{Nup-Cy} \text{,}
\end{equation}
where $\Delta \gamma=-\cos\theta_\text{Cy}$. We use the usual cylindrical polar coordinates $(r,\psi,z)$ to define positions within the pore of radius, $R$, and height, $h$, and assume axis-symmetry such that all $\psi$ dependence can be neglected. The height of the FG-Nup:cytosol interface is given by
\begin{equation}
    z_\text{Nup-Cy}(r) = \frac{(z^\star-h) (R-r)}{R-r_0}+h
\end{equation}
where $r_0=\tan(\phi/2)(z^\star-z_0)$ is the radius at which the FG-Nup condensate interface meets the conical capsid, and $z^\star$ is the height of the FG-Nup:cytosol interface where it meets the capsid. The relevant surface areas are then given by
\begin{align}
    &A_\text{Ca-Nup} = \pi  \tan (\phi/2 ) \sec (\phi/2 ) (z^\star-z_0)^2\\
    &A_\text{Nup-Cy} = \pi  \left(R^2- r_0^2\right) \sqrt{\frac{(z^\star-h)^2}{(R-r_0)^2}+1},
\end{align}
and the volume displaced by the capsid and the volume of the FG-Nup condensate for arbitrary straight line interface are given by
\begin{align}
    & V_{\mathrm{cone}} = \frac{1}{3} \pi  \tan ^2(\phi/2 ) (z^\star-z_0)^3\\
    &V_{\mathrm{Nup}} = -V_{\mathrm{cone}}+\frac{1}{3} \pi  (z^\star-h) \left(R^2+R r_0+r_0^2\right)+\pi  h R^2\text{.}
\end{align}
To enforce volume preservation of the FG-Nups we write
\begin{equation}
    V_{\mathrm{Nup}} = \pi d R^2
\end{equation}
which we then solve for $z^\star$ giving the following `physical' solution:
\begin{align}
    z^\star =& \frac{-R \cot (\phi/2 )}{4 (h-R \cot (\phi/2 )-z_0)} \Bigg[2 h+2 z_0\nonumber\\
    &+\sqrt{2} \csc (\phi/2 ) \sqrt{\cos ( \phi ) \left(3 h^2-6 h z_0+R^2+3 z_0^2\right)-3 h^2+2 R \sin ( \phi ) (h-z_0)+6 h z_0+R^2-3 z_0^2})\Bigg]\nonumber\\
    &+\frac{4 z_0 (h-z_0)+2 R^2 \cot ^2(\phi/2)}{4 (h-R \cot (\phi/2 )-z_0)}, 
\end{align}
which is shown in Fig.~\ref{fig:EmbeddingValuesSimple}. To lowest order in a series expansion in $\phi$, this is approximated by
\begin{equation}
    z^\star = h+\frac{\phi ^2 (h-z_0)^3}{4 R^2}-\frac{\phi ^3 (h-z_0)^4}{8 R^3}+O\left(\phi ^4\right).
\end{equation}
Using this solution for $z^\star$ we can proceed to calculate the energy, (Fig.~\ref{fig:phaseDiagramsSimple}). Writing $z_0=h-\epsilon$ and expanding out in $\epsilon$ we find the following approximation of the free energy around the point where the cone first enters the NPC 
\begin{equation}
    \tilde{E}=\pi  R^2+\epsilon ^2 \left(\pi  \Delta \gamma  \tan (\phi/2 ) \sec (\phi/2 )-\pi  \tan ^2(\phi/2 )\right)+O\left(\epsilon ^4\right)
\end{equation}
when the quadratic term changes sign then the capsid will wet. Thus we find the critical dimensionless surface energy density as a function of cone geometry $\phi$ is given by
\begin{equation}
    \Delta\gamma^\star = \sin(\phi/2)=\frac{\phi}{2} + O (\phi^2) = -\cos(\theta_\text{Cy})\text{,}
\end{equation}
giving a prediction for the wetting transition that agrees well with the full numerical solutions in the main text. Using the full solutions we can numerically minimise the free energy and calculate the optimal $z_0$ for a given $\Delta\gamma$ and $\phi$.

\newpage
\section{Extended Data}
Here we provide extended data to support the figures in the main paper. 

\subsection{Shape space}
We plot where different capsid shapes lie in a broader ``shape space'' by considering their volume and aspect ratio. This permits us to  identify shapes that can plausibly translocate the NPC. We do so by excluding shapes that are either too wide to fit into the central channel, sterically, or too short to bridge between the cytosol and nucleoplasm. 

\begin{figure}[!htb]
    \centering
    \includegraphics[width=\textwidth]{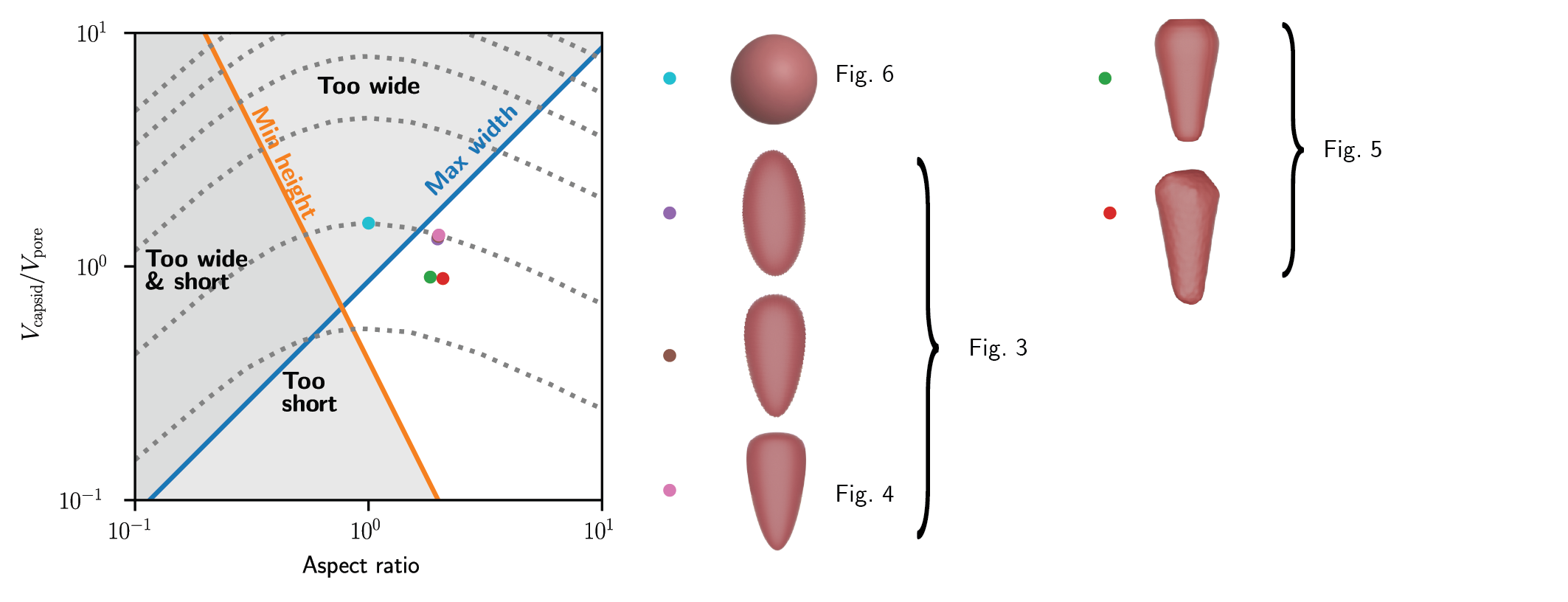}
    \caption{``Shape space'' of capsids in terms of their volume (relative to pore volume) and aspect ratio. Blue line denotes the maximum width that will fit through the pore while the orange line shows the minimum height needed to make contact with the nucleoplasm before the cytosolic side is fully enveloped. The dotted lines denote contours of fixed surface area (for purely ellipsoidal capsids, \textit{i.e.}, defined by a single aspect ratio). Points on the graph denote the shapes used in the main paper with the graphical legend showing which figures in the main paper these shapes correspond to.}
    \label{fig:S3}
\end{figure}

\subsection{Capsid reorientation}
Here we provide figures showing the full energy landscapes used to produce the minima landscape shown in Fig.~\ref{fig:3} in the main text. We also include an intermediary shape as deinfed in Table \ref{tab:params}.

\begin{table}[!htb]
\begin{tabular}{|c||c|c|c|}\hline
    Shape & $C$ & $S$ & $h$ \\ \hline
    Ellipse & $0.923077$ & $0$ & $1.84615$\\
    Half cone & $0.880038$ & $15$ & $1.84615$\\
    Cone & $0.821446$ & $21$ & $1.84615$\\ \hline
\end{tabular}
\caption{\label{tab:params}Parameter values for the $3$ simplified axisymmetric shapes defined by Eq~\ref{eq:3-parameter} with lengths in units of capsid radius}
\end{table}

\begin{figure}[!htb]
\center\includegraphics[width=\textwidth]{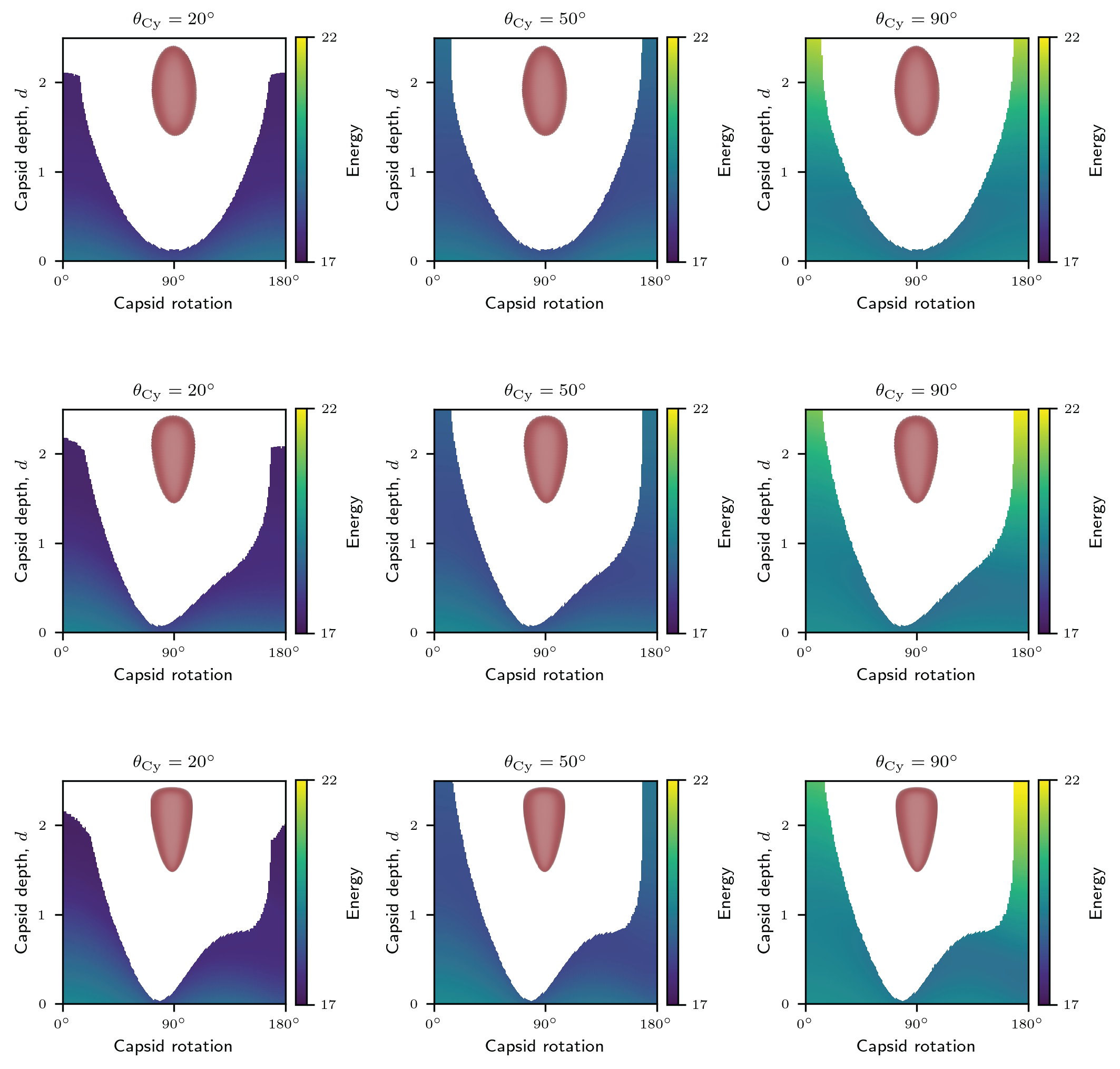}
\caption{Extended energy landscapes for all three axisymmetric capsid shapes defined by in Table~\ref{tab:params}. Capsid rotation is defined by a clockwise rotation about the capsid centre-of-mass in the plane shown. The white region missing in the centre denotes the steric boundary where the capsid makes contact with the pore edge or when it touches the nucleoplasm.}
\label{fig:S4}
\end{figure}

\begin{figure}[!htb]
    \centering
    \includegraphics[width=\textwidth]{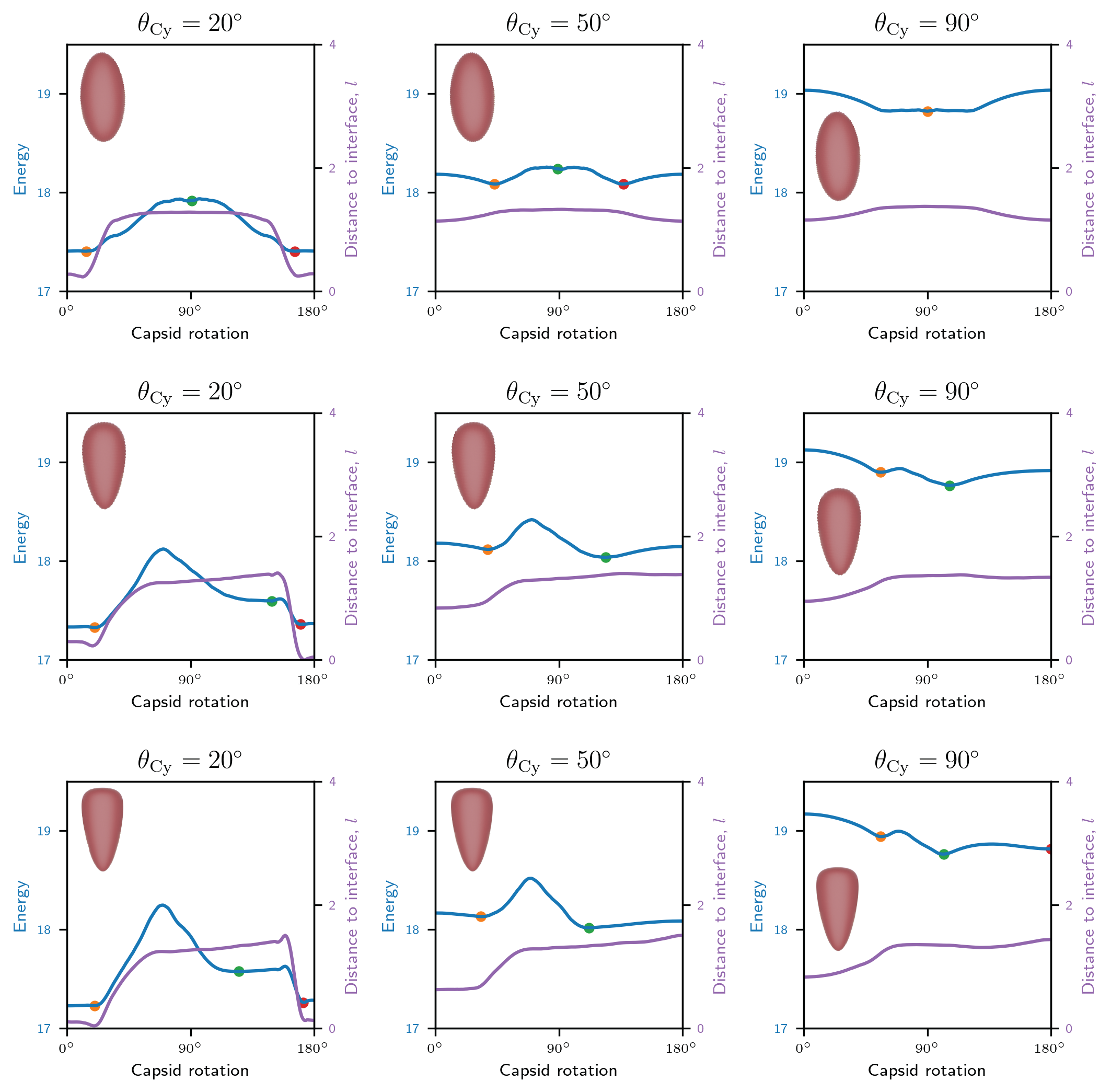}
    \caption{Minimum energy values as a function of capsid rotation for the shapes defined in Table \ref{tab:params}.}
    \label{fig:S5}
\end{figure}

\newpage
\newpage
\subsection{Translocation}
Here we include tables of energy landscapes as for a broader range of parameters along with graphs showing the areas of different interfaces as the capsids transition through the pore allowing for the snap-through definition used in the main text.

\begin{figure}[!htb]
    \centering
    \includegraphics[width=\textwidth]{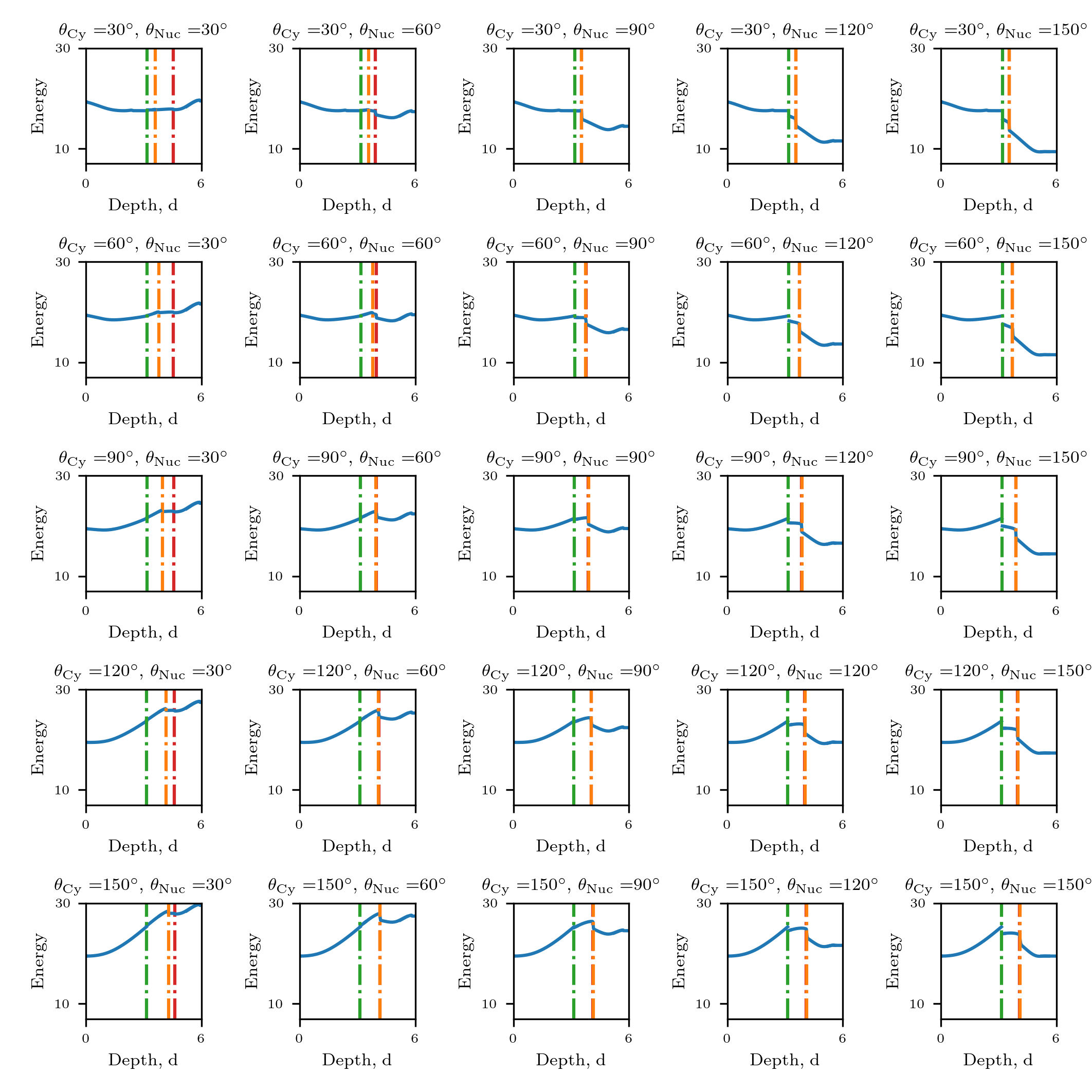}
    \caption{Energy landscapes as a function of capsid depth, $d$, for the complete capsid translocation with various values of cytoplasmic and nuclear contact angles, $\theta_\text{Cy}$ and $\theta_\text{Nuc}$. Green and orange dot-dashed lines denote the the topological transition of breaking the nuclear interface and detaching from the cytosolic interface respectively. The red dot-dashed line denotes a discontinuous change in the interfacial areas, or ``snap through'', not associated with topological changes.}
    \label{fig:S6}
\end{figure}

\begin{figure}[!htb]
    \centering
    \includegraphics[width=\textwidth]{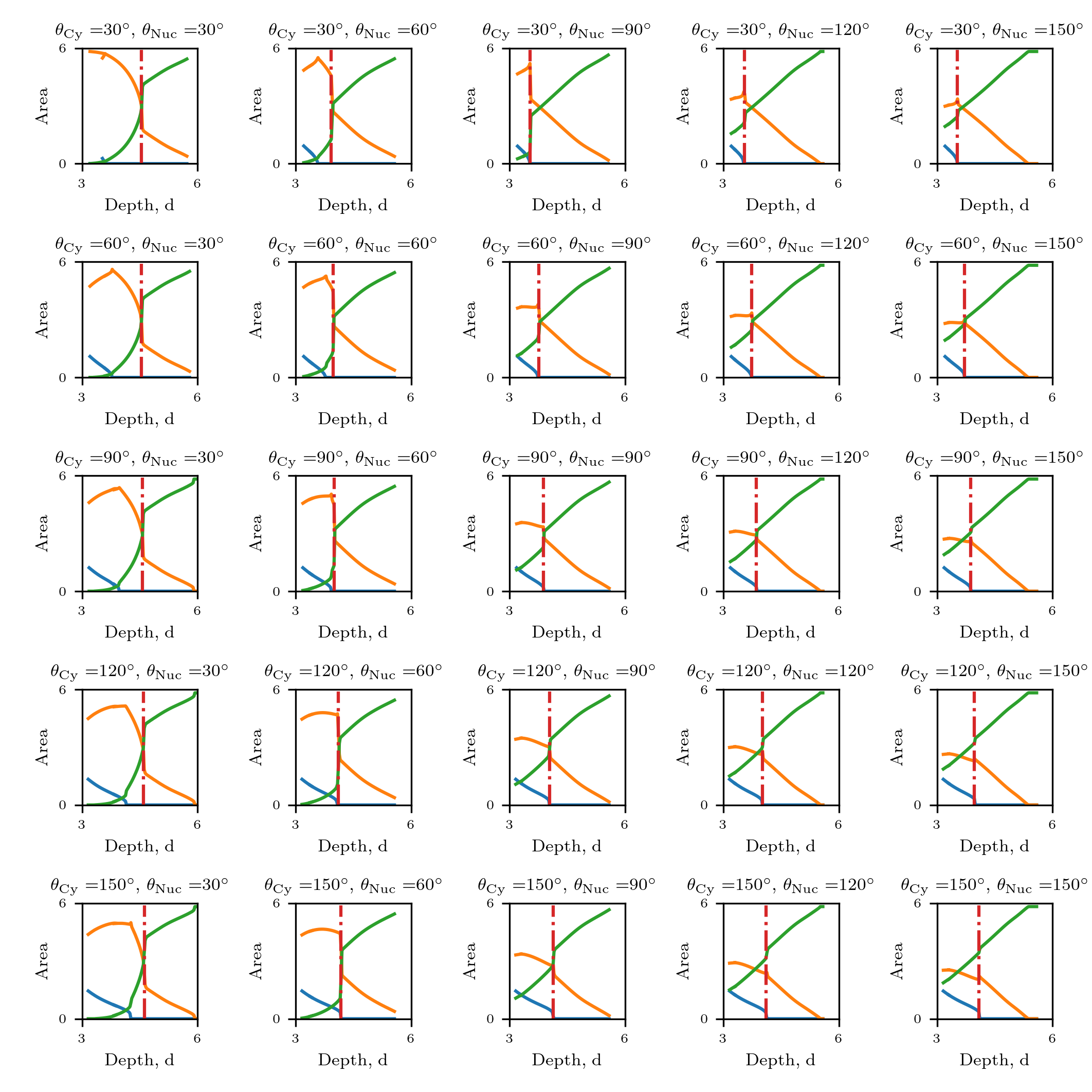}
    \caption{Areas in contact with cytosol ($A_\text{Cy}$ in blue), nuclear porin ($A_\text{Nup}$ in orange) and nucleoplasm ($A_\text{Nuc}$ in green) as a function of capsid depth, $d$, from the moment of contact with the second interface with various values of cytoplasmic and nuclear contact angles, $\theta_\text{Cy}$ and $\theta_\text{Nuc}$. The red dot-dashed line denotes a discontinuous change in the interfacial areas, or ``snap through'', not associated with topological changes.}
    \label{fig:S7}
\end{figure}

\end{document}